# Interdisciplinary Teams for Teacher Professional Development

**Carmen Fies[1] and Chris Packham[1,2]**

[1]*The University of Texas at San Antonio*

*One UTSA Circle*

*San Antonio, TX 78249*

*USA*

[2] *National Astronomical Observatory of Japan,*

*2-21-1 Osawa,*

*Mitaka,*

*Tokyo, 181-8588,*

*Japan*

**Abstract**. Secondary school teachers often lack the necessary content background in astronomy to teach such a course confidently. Our theory of change postits that an increased confidence level will increase student retention in astronomy and related STEM fields. Beyond the science content knowledge though, teachers need opportunities to embed the content in pedagogically sound practices, and with appropriate technology tools. We report on our interdisciplinary approach to designing, developing, fielding, and iteratively improving the San Antonio Teacher Training Astronomy Academy (SATTAA), an annually offered Teacher Professional Development program. In particular, we present how our separate areas of expertise, in content and in STEM pedagogy, led to a synergistic process of teacher professional development that has now resulted in three cohorts of alumni. In this paper, we share our interdisciplinary processes and lessons learned; program metrics are described elsewhere in detail.

## 1.   Background

The need for disciplinary attention to science, technology, engineering, and mathematics (STEM) Teacher Professional Development (TPD) is particularly great at a time when there are increasing trends of science-skepticism in the general population.

For example, a quick Google search for "flat earth conference" yielded 29,800 results. Although not all of these are propagating the false belief that our home planet is a flat disk, an unsettling number of adults vehemently defend this idea, or at least entertain surprising conjectures on the subject. In a similar vein, and highly problematic in the current pandemic, is the number of adults who are skeptical of vaccinations (Berman, 2020).

Although these trends initially motivated us to join in an endeavor to provide a TPD program focused on astronomy education in order to help mediate shortfalls in STEM literacy, our work was further motivated by recognizing persistent patterns of underrepresentation in STEM fields, and by troubling, and equally persistent, STEM attrition rates (e.g., Chen & Soldner, 2013).

What makes interdisciplinarity in the design of STEM TPD particularly effective is the same principle that guides teachers every day; that is, explicitly intertwine content with appropriate pedagogy and technologies. This is what drives the preparation of our program. Each design team member brings to the design and fielding of sessions areas of expertise that somewhat overlap and integrate with the areas of expertise others on the team have. By jointly developing the details of the event, each team member becomes more fluent in the disciplinary home of others; as a by-product, the professional development results in a more holistic experience for participants. In the sections below we describe the program's genesis through presenting the iterative design process.

## 2. The San Antonio Teacher Training Astronomy Academy (SATTAA)

SATTAA builds on the knowledge that children have an innate curiosity about, and fascination with, the world around them, including astronomy, rockets, dinosaurs, ancient Egypt, and similar topics of scientific interest. They want to understand how things work; yet, somewhere during their school experience many become disenchanted with learning about scientific ideas and processes.

Teachers are crucial mentors and guides to students, and are thus in a unique position to engender STEM enthusiasm into the youth. Providing high-quality TPD thus addresses a major point of leverage in contributing positively to STEM learning. In the case of astronomy education, it is troubling that most teaching of such secondary school courses is facilitated by teachers who do not have much background in astronomy. Between 2016-19, only 5.22-6.23% (Smith, 2020) of all new Texas science teacher certifications were specific to any form of physics content, which we use as proxy to astronomy content expertise. Out-of-field teaching of astronomy is the rule rather than exception in Texas schools.

SATTAA, a program for future and current STEM teachers, aims to provide opportunities to enhance content expertise towards more confident teaching of astronomy, while also providing pedagogical and technological tools and resources to make those classes fun and interesting. The goal is for SATTAA alumni to keep their students engaged with STEM learning and STEM careers. Fostering astronomy content knowledge is a core focus of the program, and is facilitated by astronomers engaged in astrophysics research and higher education. However, intertwining this content with productive pedagogies and technologies in secondary classrooms is of equal importance, and accomplished through team members with expertise in teacher preparation in general, and STEM education in particular.

In the sections that follow, we describe each iteration separately first, then summarize the experiences across the three years and include some metrics.

### 2.1. Inception: SATTAA2018 Pilot Run

In 2017, we began conversations about developing a joint summer program for future (preservice) and current (inservice) science teachers in and around San Antonio, Texas. We quickly agreed that, beyond intertwining astronomy content and STEM pedagogy, a strong program will be community-based and include partners from the informal education sector, and such a program would be offered at zero cost to participants. Supported by funding associated with NASA's Universe of Learning and administered through the Space Telescope Science Institute (STScI), and funding from the National Science Foundation (NSF), we developed a 2-week pilot program offered only to preservice STEM teachers who were enrolled in a STEM teacher preparation program at the Univeristy of Texas at San Antonio. This first iteration of SATTAA took place in the summer of 2018.

Thematically, the event was anchored in Universe of Learning resources, but each content topic was extended by astronomy Subject Matter Experts (SME), who are astronomers at the University of Texas at San Antonio (UTSA) or at the Southwest Research Institute (SwRI). In addition, the team included educators at UTSA, STScI, the McDonald Observatory[1], the Challenger Center at the Scobee Education Center[2], and the Witte Museum[3]. Through a series of conversations, the team developed a sequence of sessions that was further aligned with what faculty SMEs taught in various courses. The main adjustment to those materials consisted of embedding inquiry-based hands-on experiences. Each section was self-contained, but drew on knowledge gained in earlier sessions of the program.

As the team engaged in deeper conversations about content and pedagogical approach, we learned more about each others' fields. Coming from different academic backgrounds included speaking with different discipline-based jargon, as well as having somewhat different conceptual models of what constitutes learning. The recognition that we needed to engage in a process of calibration across the team led to initial discussions about specific content topics and potential hands-on opportunities for participants, which included Universe of Learning explorations.

The sessions with informal education partners were coordinated to seamlessly integrate with the overall sessions. All three partners already had substantive experience in providing science-related experiences in informal learning events. The McDonald Observatory and Scobee Education Center regularly provide astronomy-related public events, and the Witte Museum regularly provides a range of learning events that include science learning. The critical lens added by the Witte Museum was that of astronomy as a lived multi-cultural experience across human history.

In order to get a good understanding of the participants' views of the program, a variety of data were collected. The participants completed a pre- and post content quiz, submitted daily reflections, contributed to two focus group interviews, and agreed to recordings of their final presentations. As the focus of this paper is on the interdisciplinary team designing, developing, and fielding the summer program, only data related to participant feedback that informed iterative design decisions are included here.

---

[1] https://mcdonaldobservatory.org/
[2] http://sacscobee.org/
[3] https://www.wittemuseum.org/

Seven future teachers participated in SATTAA2018. That first iteration almost completely (90%) took place face-to-face (f2f), including all field trips (40% of the total contact time). The only exception in that year were sessions facilitated by the STScI education specialist situated in Maryland (10% of contact time). The topics ranged from exploring our solar system to cosmology, and included a specific focus on ground- and space-based telescopes. In terms of time distribution, the dominant share at 67% was dedicated to content learning; the next largest component at 20% consisted of logistics, which included travel time associated with field trips. The remaining 13% of time was dedicated to content-specific pedagogy and participant presentations.

From the instructional team's point-of-view, the program rolled out smoothly. Nearly all of the logistics arrangements worked well, and session content flowed logically from day-to-day. However, in this pilot run, not all sessions had a hands-on inquiry component and the specific details were not sufficiently closely coordinated across sessions. In addition, the time expectations for the field trip to the McDonald Observatory were overly demanding. On those three days, team facilitators and participants dedicated time to the program from pre-dawn until past midnight; this includes night-time sky observations, as well as the drive to and from the site. Although every moment on site had been exciting to the 2018 participants, ultimately, all were exhausted.

The participants' point-of-view, given in the form of oral and written feedback, although largely positive, afforded us specific action items for the upcoming cycle. Particularly encouraging was that all seven participants indicated they not only appreciated the experience as such, but also had a better understanding of astronomy now and would recommend the program to others. Six of the participants indicated that the distribution of time, in terms of f2f vs. online modalities, worked well for them. The specific recommendations for improvement consisted of (1) generally revising the session format as too much of the on-campus content meetings took the shape of lectures; (2) revisiting content across the two weeks as there were some redundancies across sessions; and (3) reconsidering the time table as the program was too demanding in terms of contact hours.

### 2.2. Second iteration: SATTAA2019

The entire 2018 SME team was invited to participate in the redesign of the content sessions, and nearly all engaged in the process as they planned to facilitate sessions again in 2019. In addition, our informal education partners participated in reshaping field trips. The first cycle of revision focused on the participants' recommendations to shift sessions further away from a lecture model and to avoid existing content redundancies. We began by reviewing the content of each of the 2018 sessions to identify such redundancies, and by tracing the flow of the program in terms of logistics. At the core of many discussions thus were considerations of more interactive engagement and a better streamlined sequence. At the same time, and although not recommended by 2018 participants, the team paid attention to connecting program experiences more explicitly to secondary school curricula.

The redesign led to a higher number of opportunities for hands-on exploration of astronomy content, and to a more streamlined program. Through avoiding redundancies, it also was possible to add a topic: terrestrial climate change. In addition, more time could be dedicated for participants to develop pedagogical connections to content. Specifically, in addition to spending time on working with pre-existing secondary lesson plans, the participants created their own lessons in teams. The field trips also were redesigned; most markedly in reducing the hours of engagement associated with the field trip to the observatory.

In 2019, we broadened participation by opening registration to current and future teachers. This change enabled us to include collaborative opportunities for those who were preparing to teach, with those already experienced in leading science lessons in school classrooms. Our expectation was that collaborations that intentionally teamed representatives from both groups would be beneficial for two reasons: (1) Future teachers could learn from the lived experiences of current teachers; and (2) current teachers would be able to take a fresh look at the teaching profession through the eyes of their preservice counterparts. Because Texas teachers must earn credits each year to maintain their status as certified teachers, we added state-approved Continuing Professional Education (CPE) credits that inservice participants earned through participation in the program.

All measurements from 2018 were kept intact to have a point of comparison between the two years. However, since inservice teachers were added to the participant pool, additional probes were necessary to explore the effects of pairing future with current teachers in teams. In particular, we sought to learn what the participants' perceptions of the experience were, and what they thought might have been different for them had the groupings not been mixed.

SATTAA2019 was conceptualized and offered as primarily a f2f program, including all field trips. However, all afternoons not associated with field trips took place online. The distribution of time overall was similar to the prior year: at 63%, the majority of time was dedicated to content learning. No improvement regarding time for logistics could be made (21%), however, as in the year before, this included travel time associated with field trips. The remaining 16% of time were dedicated to content-specific pedagogy and participant presentations.

A total of six pre- and five inservice teachers participated. We learned from feedback to the 2019 iteration that, as in the pilot, all participants appreciated the experience, had a better understanding of astronomy, and would recommend the program to others. Although the distribution of contact hours between f2f and online engagement changed between the two iterations, the participants expressed satisfaction with logistics in both cases. This may be indicative of teachers' primary concern with content and goals, and a willingness to adapt to logistics as long as the program captivates their interest.

All participants stated that they appreciated being part of mixed teams consisting of pre- and inservice teachers. However, inservice teachers also noted that there would have been benefits in being grouped with other inservice teachers. Specifically, they noted that they would have liked to have used that time to exchange tips of what works in their respective classrooms. They also would have liked to co-plan lessons for the upcoming academic year and then stay in contact with the team partner during lesson implementation to compare their experiences and further refine those lessons.

The main recommendation for improvement was to create a website associated with the program. The participants noted that staying connected beyond the 2-week program was important to them and a website could serve as a sustained networking hub for them.

### 2.3. Third iteration: SATTAA2020

The COVID-19 pandemic drove a dramatic shift for the program: the mainly f2f approach had to be transformed quickly and completely to online interactions. This change drove all other changes in preparing SATTAA2020.

A primary benefit of the fully online format was that we were able to further diversify our team of SME presenters by including astronomers located elsewhere in the USA. We paid special attention to ensuring a strong representation of female astronomers to present the materails to ensure teachers experience gender diversity in the field. For details please see Table 1 in Section 2.4. However, as is reflective of the situation in astronomy in general, the racial diversity is not as one would wish for, and this is an issue we hope to address in future SATTAA iterations. However, our program features strong multi-cultural components, discussing, for example, the aboriginal people of Texas' interpretation of the night sky, and using local cave paintings of the night sky as our bridge between culture and astronomy.

Again, the range of topics was broadened, this time adding a new session specific to solar eclipses, and a focus on the role of coronagraphs in finding exoplanets. In week 2, participants joined an online introductory astronomy class for four class sessions. Field trips also moved to virtual modalities; we were able to keep two of the three prior opportunities, but could not equivalently replace the trip to the McDonald Observatory with an experience that would have allowed night-sky viewing in a dark skies setting. This was a tremendous loss of what participants' in the prior two years had deemed to be a highlight. However, by replacing the observatory field trip with another virtual option, we were able to drastically reduce logistics time (only 4% in 2020). As a result, a larger component could be dedicated to content learning (72%), and to content-specific pedagogy and participant presentations (24%). In 2020, 15 inservice and 4 preservice teachers participated in the 2-week program.

The urgent shift to a fully online program benefited from the interdisciplinarity of the team by the network connections of the astronomers, and the online pedagogical experience of the STEM educator. From the instructor point-of-view, several aspects were succeeded: Zoom as a virtual meeting platform proved to work well; the very few technical issues that occurred were associated with Internet speed and could be resolved quickly. The SATTAA2020 program was distributed ahead of time and included zoom links to all sessions, making it easy for everyone to connect. The program also included lists of materials for hands-on experiences that were part of sessions, making it possible for participants to have the materials on hand and joining as they wished. All sessions were recorded and made available online for later viewing by the attendees, but they have been rarely watched. Because of missing the physical presence in the same space, what would have been collaborative inquiry lacked the joint engagement with the same physical objects.

The feedback from participants was encouraging as it matched that received in the prior years of the program. Again, all participants appreciated the experience, indicated they had a better understanding of astronomy, and noted that they would recommend the program to others. That this outcome was identical, can be interpreted as an endorsement of the online format. What was particularly positive was that participants noted how they benefited from the online format; we even learned that some of the participants could not have joined the program if it had not been fully online. Thus the reach of SATTAA in an online environment is improved, and can assist in further diversifying our audience. For example, rural areas often are in particular need of high quality TPD programs. Although the participants felt that the virtual field trips were enriching, they noted that they wished they could have had the f2f experience.

In addition to finding a way to support f2f field trips, The participants recommended to further increase the time spent on pedagogical connections; all expressed how beneficial the opportunity was as it also served to deepen the CoP forming as part of the program. While schools are in session, teachers rarely have time to engage in deep lesson planning with peers.

In 2020, we had the opportunity to solicit competive grant applications from SATTAA Alumni to further advance astronomy teaching. We had 3 tiers of grant levels available ($2,500, $1,000, and $500), and we were substantially over-subscribed. In their applications, the teachers had to include the following items, which formed the basis of our grading metric (all equally important):

    1.How this would be used in class?
    2.The need for this project in class
    3.Comments about why this resource has not been provided by other routes
    4.Connection to SATTAA
    5.Demographics of the community served

The funded projects ranged from the construction of Mars rover models, to moon phase t-shirts, and to telescopes. We plan to run this opportunity on an annual basis, where the funds were generously provided by a local business (see acknowledegments below). This will be one futher way to support an inclusive and diverse SATTAA alumni network, which we believe will help to share resources and foster cross-school collabrations.

**2.4. Across the three years**

This section serves to provide several data snapshots across the three years. We begin with an overview of the demographics of the instructional team, of the participants, and of the inservice teachers' school demographics.

Although the core team remains the same, our overall team make-up changes each year, in part due to the natural movements of individuals shifting into other professional roles, and in part through changing demands on time. We present here a demographic summary of the instructional team make-up in terms of gender, and role in relation to science education (Table 1).

Table 1. Instructional team demographics by year, gender, and education sector

| Gender and employment category | 2018 | 2019 | 2020 |
|---|---|---|---|
| Female, total | 7 | 5 | 10 |
| *Higher education and research* | *(4)* | *(4)* | *(6)* |
| *Informal education* | *(3)* | *(1)* | *(4)* |
| Male, total | 8 | 6 | 6 |
| *Higher education and research* | *(6)* | *(4)* | *(3)* |
| *Informal education* | *(2)* | *(2)* | *(3)* |
| Total | 15 | 11 | 16 |

We are casting these demographics against those of the participating teachers (Table 2). Not by design, but as an interesting coincidence, the changing gender ratios of facilitators and participants show similar trends across the three years. Also worth noting is that in 2018, program facilitators soundly outnumbered participants, while matching exactly in 2019, and finally participants outnumbering facilitators in 2020.

Table 2.  Participant demographics by year, gender, and teaching status

| Gender and teaching status | | 2018 | 2019 | 2020 |
|---|---|---|---|---|
| **Female, total** | | 3 | 5 | 15 |
| | *Inservice teachers* | *(0)* | *(3)* | *(11)* |
| | *Preservice teachers* | *(3)* | *(2)* | *(4)* |
| **Male, total** | | 4 | 6 | 4 |
| | *Inservice teachers* | *(0)* | *(2)* | *(4)* |
| | *Preservice teachers* | *(4)* | *(4)* | *(0)* |
| Total | | 7 | 11 | 19 |

All inservice teachers worked at different area schools and, notably, at schools with a higher proportion of underrepresented minority students, and with higher poverty rates, than city-wide population averages (Table 3). To provide context, based on the U.S. Census Bureau (2020), as of July 1, 2019 San Antonio is home to roughly 1.5M inhabitants. The population majority belongs to minority demographics; here, 'white alone, not Hispanic or Latino' accounts for 24.8% of the overall population, while Hispanic or Latino is the largest component at 64.2%. Citywide, 18.6% of the population live in poverty. *Our program strives to reach underrepresented minority learners and the data bear out that we are successful in that endeavor.*

Table 3: In-service teachers' school demographics (percent averages).

| Year | Students: % White | Students: % Economically Disadvantaged | Students: % English Learners | Students: % Special Education |
|---|---|---|---|---|
| 2019 | 15.3 | 55.9 | 10.0 | 11.3 |
| 2020 | 11.4 | 62.9 | 9.8 | 10.0 |

A valid question is whether the program improves the understanding of astronomy amongst participants. Although this is discussed elsewhere, we offer a data snapshot here. At this point, we have pre- and posttest data of content knowledge of various astronomical phenomena for 18 of the participants. In this sample, knowledge, as measured by correct responses to 15 moderately difficult questions, improved between pretest ($M = 6.83$, $SD = 3.167$) and posttest ($M = 8.72$, $SD = 2.845$). The results of a t-test for paired samples indicate a statistically highly significant gain ($t = 5.24$; $n = 18$; $p = 0.000$) in learning outcomes[4]. The effect size of 1.24, which means that posttest scores are over one standard deviation better than the pretest scores, is considered high.

Twice in each SATTAA iteration, once at the beginning and again at the end of the program, we asked participants to tell us, on a 5-point scale, how interested they were in, and how much they felt they understood of, the astronomy content explored during SATTAA. In all cases, the results indicate a positive change (Table 4).

---

[4] M = mean value; SD = standard deviation; t = t-value, the size of difference relative to variation; p = probability that results are a chance occurrence

Table 4: Means of pre- and post results of survey data, rounded (Likert Scale: 5 = very high, 1 = not at all)

| Year | Status | n | Interest | | | Understanding | | |
|---|---|---|---|---|---|---|---|---|
| | | | pre | post | change | pre | post | change |
| 2018 | Preservice | 7 | 3.6 | 4.7 | **+1.1** | 3.4 | 4.1 | +0.7 |
| 2019 | Preservice | 6 | 4.8 | 5 | +0.2 | 3.7 | 4.5 | +0.8 |
| 2019 | Inservice | 5 | 4 | 4.8 | +0.8 | 2.8 | 4.3 | +1.5 |
| 2020 | Preservice | 4 | 4 | 5 | **+1.0** | 2.5 | 4.5 | **+2.0** |
| 2020 | Inservice | 15 | 4.3 | 4.9 | +0.6 | 3.4 | 4.6 | +1.2 |

Although the biggest gains in self-reported interest are amongst preservice teachers, as is the single biggest gain in understanding, these data alone tell us little about the participants' dedication to and confidence in providing astronomy education. Comparing the means of the same data across the three years, we see a positive trend for self-reported changes in the understanding of content knowledge. The associated pattern for changes in interest is more varied, but also shows that each cycle resulted in a positive change (Figure 1).

Figure 1. Means of pre- and post results of survey data specific to interest and understanding of content knowledge, by year.

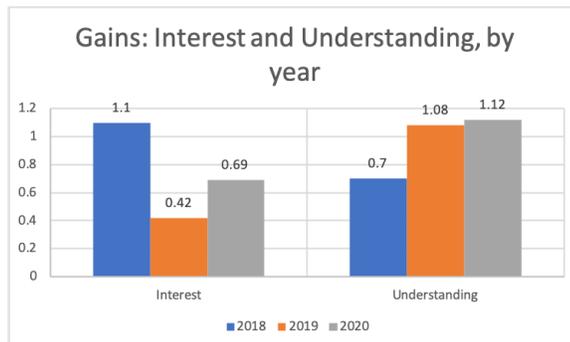

In summary, we see positive trends. Although, the numbers are still to small to execute a robust quantitative analysis, when combined with the qualitative data collected, there is strong indication that participants feel not only better prepared to teach astronomy in their own classrooms, but also feel enthusiastic about it.

## 3. The Value of Interdisciplinary Teams for STEM TPD

In a recent publication by Stephen Pompea and Pedro Russo (2020), the authors describe the current landscape of astronomy education and conclude that effective efforts benefit from intertwining the knowledge of scientific content with appropriate pedagogical practices. Across the three years of SATTAA, the iterative design process, guided by feedback from instructors and participants, brought to the fore a need for continued calibration around the learning and teaching of astronomy content.

The participating teachers, pre- and inservice alike, consistently have noted they felt enriched in content knowledge, in pedagogical connections, as well as in net-working with facilitators and with peers. To help pariticants remain connected beyond the program, we established the SATTAA Alumni group that consists of all participants who completed the program to date. This extends the network in concrete and sustained means into a Community of Practice (CoP). The power of a CoP (Wenger, 1998) lies in the shared disciplinary goals of participants. In essence, they enter a mutually beneficial relationship around a shared practice in which each participant has legitimate ways to contribute and to become more expert in the process. The SATTAA Alumni group has at least one weekly touch-point in the form of Astro Snacks, a listserv that provides links to astronomy education news and resources.

Much like SATTAA participants have become part of a CoP, the instructional team has become a CoP through engaging iteratively in the work of offering an astronomy education TPD that is soundly grounded in the theory and practice of astronomy, as well as in the theory and practice of STEM education. A critical advantage of SATTAA in the online format is the ability to invite SMEs from far away, leveraging existing collaborations and improving team diversity. Further, we note the reduced environmental footprint that an on-line meeting has, as discussed in Burtscher et al. (2020).

**Acknowledgements**. We thank NASA, the Space Telescope Science Institute, and NSF grant no. 1616828 for supporting this work financially. We also gratefully acknowledge the generous support of BEAT LLC (https://beatllc.com) for providing the assistance to enable SATTAA grants starting 2020.